\documentclass[11 pt]{article}
\usepackage[mathscr]{eucal}
\usepackage{amssymb,latexsym}
\usepackage{verbatim}
\usepackage{amsmath}
\usepackage{amsthm}
\usepackage{enumerate}
\usepackage{authblk}
\usepackage{color}
\usepackage{multicol}
\usepackage{tikz}
\usepackage{xypic}
\usepackage{caption}
\usepackage{cite}
\usepackage{hyperref}
\usetikzlibrary{snakes}

\usepackage[nottoc,notlot,notlof]{tocbibind}

\newtheorem{thm}{Theorem}[section]
\newtheorem{lem}[thm]{Lemma}
\newtheorem{prop}[thm]{Proposition}

\hypersetup{
    colorlinks=true,
    linkcolor=blue,
    filecolor=magenta,      
    urlcolor=cyan,
    citecolor=black,
}

\newcommand\wt{\widetilde}

\renewcommand\S{\Sigma}

\newcommand\s{\sigma}

\newcommand\e{\varepsilon}

\newcommand\g{\gamma}

\renewcommand\a{\alpha}

\newcommand\beq{\begin{equation}}
\newcommand\eeq{\end{equation}}
\newcommand\ben{\begin{enumerate}}
\newcommand\een{\end{enumerate}}
\newcommand\bit{\begin{itemize}}
\newcommand\eit{\end{itemize}}

\newcommand{\R}{\mathbb R}

\newcommand{\pd}{\partial}

\def\undertilde#1{\mathord{\vtop{\ialign{##\crcr
   $\hfil\displaystyle{#1}\hfil$\crcr\noalign{\kern1.5pt\nointerlineskip}
   $\hfil\tilde{}\hfil$\crcr\noalign{\kern1.5pt}}}}}

\newcounter{mnotecount}

\title{The $C^0$-inextendibility of some spatially flat FLRW spacetimes}

\author{Eric Ling\footnote{eric.ling@univie.ac.at}}

\affil{University of Vienna, Faculty of Mathematics
\linebreak
Oskar-Morgenstern-Platz 1, A-1090 Wien, Austria}

\begin{document}
\date{}
\maketitle
\vspace{.0in}

\begin{abstract}
Utilizing some of Sbierski's recent $C^0$-inextendibility techniques \cite{Sbierski_C0_FLRW}, we prove the $C^0$-inextendibility of a class of spatially flat FLRW spacetimes without particle horizons.
\end{abstract}




\section{Introduction}

In \cite{Sbierski_C0_FLRW}, Sbierski proved the $C^0$-inextendibility of a class of FLRW spacetimes without particle horizons. His main results, Theorems 1.5 and 1.6, apply to FLRW spacetimes which are spatially spherical and spatially hyperbolic, respectively. In this paper, we complement his results by establishing the $C^0$-inextendibility of a class of spatially flat FLRW spacetimes, also without particle horizons. As in \cite{Sbierski_C0_FLRW}, we work in the class of simply connected models,
\[
M = I \times \R^n \quad \text{ and } \quad g = -dt^2 + a(t)^2\big((d u^1)^2 + \dotsb + (d u^n)^2\big),
\]
where $I$ is an open interval and the warping function $a \colon I \to (0,\infty)$ is called the \emph{scale factor}.

Our definitions, notations, and conventions follow \cite{Sbierski_C0_FLRW} which we briefly review. A $C^k$ Lorentzian manifold $(M,g)$ is one where the metric $g$ is of regularity class $C^k$. Regardless of the regularity class of the metric, the manifold $M$ is always assumed be smooth. A $C^k$ \emph{spacetime} is a time-oriented $C^k$ Lorentzian manifold $(M,g)$ which is connected. 

 A $C^0$ \emph{extension} of a smooth spacetime $(M,g)$ is an isometric embedding \[\iota \colon (M,g) \hookrightarrow (\widetilde{M}, \wt{g}),\] where $(\wt{M}, \wt{g})$ is a connected $C^0$ Lorentzian manifold with the same dimension as $M$ and $\iota(M) \subsetneq \widetilde{M}$ is a proper open subset. Given a $C^0$ extension, the \emph{past boundary} $\partial^-\iota(M)$ is the subset of the topological boundary $\partial \iota(M)$ given by the set of points $p \in \widetilde{M}\setminus \iota(M)$ such that there is a $C^1$ timelike curve $\gamma \colon [a,b] \to \widetilde{M}$ such that $\gamma(a) = p$ and $\gamma\big((a,b]\big) \subset \iota(M)$ and $\iota^{-1}\circ \gamma|_{(a,b]}$ is future-directed in $M$. The \emph{future boundary} $\partial^+\iota(M)$ is defined time-dually.
 
If no $C^0$ extension of a smooth spacetime $(M,g)$ exists, then we say $(M,g)$ is $C^0$\emph{-inextendible.} It's known that if a $C^0$ extension of a smooth spacetime $(M,g)$ does exist, then either the future or past boundary is nonempty. We say $(M,g)$ is \emph{past} $C^0$\emph{-inextendible} if the past boundary of any $C^0$ extension is empty. 

Our main theorem is the following.

\begin{thm}\label{thm: main}
Let $(M,g)$ be a spatially flat  simply connected FLRW spacetime with \linebreak dimension $n+1\geq 3$. Suppose the scale factor $a(t)$ is smooth and satisfies the following.
\begin{itemize}
\item[\emph{(i)}] $a(t)$ is defined for all $t \in \R$ and $a(t) \to 0$ as $t \to -\infty$,
\item[\emph{(ii)}] $\int_{t}^1\frac{1}{a(s)}ds \to \infty$ as $t \to -\infty$,
\item[\emph{(iii)}] $a(t)\int_t^1 \frac{1}{a(s)}ds \to \infty$ as $t \to -\infty$.
\end{itemize}
Then $(M,g)$ is past $C^0$-inextendible.
\end{thm}

\medskip

\noindent\emph{Remarks.}\:

\begin{itemize}
 \item[(1)] If the scale factor satisfies (i), then $t = -\infty$ is sometimes referred to as the ``big bang." 
 
 \item[(2)] The theorem generalizes to the case that $a(t)$ is only defined on an interval $I = (-\infty, t_f)$ for some $t_f \in \R$; the proof is the same. These spacetimes are called ``past eternal" since $t \to -\infty$. 
 
 \item[(3)] If the scale factor satisfies (ii), then $(M,g)$ is said to ``not possess particle horizons." 
 
 \item[(4)] Condition (ii) in Theorem \ref{thm: main} is redundant since it's implied by (i) and (iii). 
 
 \item[(5)] If the scale factor satisfies (i) and (ii), but, instead of (iii), it satisfies $a(t)\int_t^1\frac{1}{a(s)}ds \to c \in (0,\infty)$ as $t \to -\infty$, then $(M,g)$ is known to be past $C^0$-extendible \cite{GLQ}. For example, take $a(t) = e^t$. 
 
 \item[(6)] If $(M,g)$ in Theorem \ref{thm: main} is also future complete, then it's known that it's also future $C^0$-inextendible \cite{GLS}; therefore $(M,g)$ is $C^0$-inextendible in this case.
\end{itemize}

\medskip

\noindent\emph{Example.} Consider $a(t) = e^{-\sqrt{-t}}$ for $t \leq -1$, arbitrarily extended to a smooth positive function for $t > -1$. Then $a(t)$ satisfies (i) - (iii) of Theorem \ref{thm: main}. Therefore the corresponding spatially flat simply connected FLRW spacetime $(M,g)$ is past $C^0$-inextendible. Note that, at first glance, it might seem like $(M,g)$ is past timelike complete since $t$ can limit to $-\infty$. If that were the case, then the main result in \cite{GLS} would already suffice to prove past $C^0$-inextendibility of $(M,g)$, rendering Theorem \ref{thm: main} redundant. However, $(M,g)$ is \emph{not} past timelike complete. While the vertical timelike geodesic $t \mapsto (t, u^1_0, \dotsc, u^n_0)$ is past complete for any fixed point $(u^1_0, \dotsc, u^n_0)$ in $\R^n$, any other timelike geodesic---specifically those with an initial tangent vector not parallel to $-\partial_t$---will be past incomplete.

\medskip

The investigation of low-regularity inextendibility results is motivated in part by a desire to understand and classify the strength of gravitational singularities. See the nice discussion in \cite{Sbierski_holonomy}. This investigation began with Sbierski's seminal work \cite{Sbierski_C0_original} on the $C^0$-inextendibility of the positive-mass Schwarzschild spacetime\footnote{The negative-mass Schwarzschild spacetime is also $C^0$-inextendible which follows from \cite{Minguzzi_Suhr} since the spacetime is timelike complete. This answers one of the questions raised in \cite{chrusciel2021quo}.}. Since then other results have been found \cite{Sbierski_2018, Sbierski_holonomy, Gal_Ling_C0_remarks, Sbierski_C0_FLRW, GLS, Graf_Ling, Minguzzi_Suhr, annoying_null_boundaries, Grant_2018, SbierskiUniqueness, Graf_Beld-Serrano, Miethke, Sbierski_inventiones, Le1, Le2}. Nevertheless, the understanding of low-regularity inextendibility is far from complete.

\section{The proof}

\subsection{Characterization of TIFs}

In this section, we characterize the TIFs of the spatially flat FLRW spacetimes appearing in Theorem \ref{thm: main}. Along the way, we will prove the $C^0$-extendibility result mentioned in remark (5) after Theorem \ref{thm: main}.

Consider the spatially flat FLRW spacetime $(\hat{M}, \hat{g})$ written out in comoving coordinates:
\[
\hat{M} \,=\, \R \times \R^n \quad \text{ and } \quad \hat{g} \,=\, -d\hat{t}^2 + e^{2 \hat{t}}\big((d u^1)^2 + \dotsb + (du^n)^2\big).
\]
Introducing spherical normal coordinates $(\hat{t}, r, \omega)$ centered at the origin $u^i = 0$, we can rewrite the metric as 
\[
\hat{g} \,=\, -d\hat{t}^2 + e^{2\hat{t}}(dr^2 + r^2 d\Omega^2),
\]
where $d\Omega^2$ is the round metric on the unit sphere $\mathbb{S}^{n-1}$.

We introduce new spherical coordinates $(T,R,\omega)$ defined via
\begin{equation}\label{eq: T R}
e^{\hat{t}} \,=\, \frac{\sin T + \cos R}{\cos T} \quad \text{ and } \quad r \,=\, \frac{\sin R}{\sin T + \cos R}.
\end{equation}
The spherical coordinates $(T,R,\omega)$ produce the following diffeomorphism
\[
\R \times (0,\infty) \times \mathbb{S}^{n-1} \ni (\hat{t}, r, \omega) \quad \mapsto \quad \big(T(\hat{t},r),R(\hat{t},r),\omega\big) \in D,\]
where
\begin{equation}\label{eq: U}
D := \big\{(T,R) \mid T \in (-\tfrac{\pi}{2}, \tfrac{\pi}{2}),\, R \in (0,\pi),\, T > R - \tfrac{\pi}{2} \big\} \times \mathbb{S}^{n-1}. 
\end{equation}
The metric in $(T,R,\omega)$-coordinates is 
\[
\hat{g} \,=\, \frac{1}{\cos^2 T}(-dT^2 + dR^2 + \sin^2R\, d\Omega^2).
\]
Therefore $(\hat{M}, \hat{g})$ is conformal to the open subset $D$ of the Einstein static universe, save for the $\hat{t}$-line at $r=0$ which is in one-to-one correspondence with the $T$-line at $R = 0$ along $-\tfrac{\pi}{2} < T < \tfrac{\pi}{2}$  within the Einstein static universe. 

Now let $(M,g)$ be as in Theorem \ref{thm: main} so that
\[
M \,=\, \R \times \R^n \quad \text{ and } \quad g \,=\,-dt^2 + a(t)^2\delta,
\]
where $\delta$ is the Euclidean metric on $\R^n$.

Define
\begin{equation}\label{eq: t-hat}
\hat{t}(t) \,=\, -\ln \left(\int_t^1 \frac{1}{a(s)}ds \right).
\end{equation}
Then $\hat{t}$ maps $(-\infty, 1)$ diffeomorphically onto $\R$ and the metric $g$ is conformal to $\hat{g}$:
\[
g \,=\, -dt^2 + a(t)^2\delta \,=\, \left(a\big(t(\hat{t})\big)\int_{t(\hat{t})}^1\frac{1}{a(s)}ds\right)^2\big(-d\hat{t}^2 + e^{2\hat{t}}\delta \big).
\]

Define 
\begin{equation}\label{eq: M_1}
M_1 \,=\, (-\infty, 1) \times \R^n \,\subset\, M \quad \text{ and } \quad g_1 \,=\, g|_{M_1}.
\end{equation}
Then $(M_1, g_1)$ is conformal to $(\hat{M}, \hat{g})$ and hence conformal to the open subset $D$, defined by \eqref{eq: U},  of the Einstein static universe (modulo the trivial coordinate singularity along $r = 0$).

\medskip 

\noindent\emph{Remark.} We verify the $C^0$-extendibility result mentioned in remark (5) after Theorem \ref{thm: main}. Suppose $(M,g)$ satisfies the hypotheses of Theorem \ref{thm: main}, except that,  instead of condition (iii), we have $a(t)\int_t^1\frac{ds}{a(s)} \to c \in (0,\infty)$ as $t \to -\infty$. Then the metric is nondegenerate along the part of the boundary of $D$ given by $T = R - \tfrac{\pi}{2}$ for $-\tfrac{\pi}{2} < T < \tfrac{\pi}{2}$ which yields a past $C^0$ extension for $(M_1, g_1)$ and hence one for $(M,g)$ as well.

\medskip

The following proposition characterizes the TIFs in $(M_1, g_1)$. It's an adaptation of \cite[Prop. 3.8]{Sbierski_C0_FLRW} to the spatially flat setting.

\smallskip

\begin{prop}\label{prop: TIFs}
Let $(M,g)$ be as in Theorem \emph{\ref{thm: main}} and define $(M_1, g_1)$ by \eqref{eq: M_1}.   Let $(T,R,\omega)$ be the coordinates on $M_1$ defined by \eqref{eq: T R} and \eqref{eq: t-hat}.
\begin{itemize}
\item[\emph{(a)}] Let $\g$ be a future-directed past-inextendible timelike curve in $M_1$  parameterized by $t$.
\begin{itemize}
\item[\emph{(i)}] The limit $T_* : = \lim_{t \to -\infty}T\circ \g(t)$ exists and $T_* \in [-\tfrac{\pi}{2}, \tfrac{\pi}{2})$.
\item[\emph{(ii)}] If $T_* = -\tfrac{\pi}{2}$, then $\bigcup_{-\infty < t} I^+\big(\g(t), M_1\big) = M_1$.
\item[\emph{(iii)}] If $T_* \in (-\tfrac{\pi}{2}, \tfrac{\pi}{2})$ , then $\omega \circ \g(t)$ converges in $\mathbb{S}^{n-1}$ as $t \to -\infty$. 
\end{itemize}

\item[\emph{(b)}] Let $\g_1$ and $\g_2$ be two future-directed past-inextendible timelike curves in $M_1$ both parameterized by $t$. Set $T_{i} = \lim_{t \to -\infty} T\circ \g_i(t).$
 Assume $T_{i} \in (-\tfrac{\pi}{2}, \tfrac{\pi}{2})$ for $i = 1,2$, and let $\omega_{i} \in \mathbb{S}^{n-1}$ denote the limit of $\omega\circ\g_i(t)$ as $t \to -\infty$. 
\begin{itemize}
\item[\emph{(i)}] If $T_1 = T_2$ and $\omega_1 = \omega_2$, then $\bigcup_{-\infty < t} I^+\big(\g_1(t), M_1\big) = \bigcup_{-\infty < t} I^+\big(\g_2(t), M_1\big)$.
\item[\emph{(ii)}] If $T_1 < T_2$ and $\omega_1 = \omega_2$, then $\bigcup_{-\infty < t} I^+\big(\g_1(t), M_1\big) \supsetneq \bigcup_{-\infty < t} I^+\big(\g_2(t), M_1\big)$.
\item[\emph{(iii)}] If $\omega_1 \neq \omega_2$, then $\bigcup_{-\infty < t} I^+\big(\g_1(t), M_1\big) \neq \bigcup_{-\infty < t} I^+\big(\g_2(t), M_1\big)$.
\end{itemize}
\end{itemize}
\end{prop}

\proof\,

\begin{itemize}
\item[(a)] 
\begin{itemize}
\item[(i)] This is clear since $T$ is a time function. 
\item[(ii)] Define $u = T + R$. Let $\nabla $ denote the gradient with respect to the metric for the Einstein static universe. Then $\nabla u = -\pd_T + \pd_R$ is past-directed null. Hence $u\circ \g(t)$ is a strictly increasing function. Therefore $u\circ \g(t)$ converges as $t \to -\infty$ and hence so does $R\circ \g(t)$. Let $u_*$ and $R_*$ denote their limits. Since $\g$ is past-inextendible within $M_1$, it follows that $R_* = T_* + \tfrac{\pi}{2}$ and hence $u_* = 2T_* + \tfrac{\pi}{2}$. If $T_* = -\tfrac{\pi}{2}$, then $R_* = 0$ and the conclusion follows from the geometry of the Einstein static universe.
\item[(iii)] Assume $T_* \in (-\tfrac{\pi}{2},\tfrac{\pi}{2})$. Reparameterize $\g$ by $T$ and represent $\g(T)$ by the curve $T \mapsto \big(T, R(T), \omega(T)\big)$. Since $\g$ is timelike, we have 
\[
0 \,>\, -1 + \left(\frac{dR}{dT}\right)^2 + \sin^2\big(R(T)\big)|\omega'(T)|^2_{\mathbb{S}^{n-1}}.
\]
 Therefore
\[
|\omega'(T)|_{\mathbb{S}^{n-1}} \,<\, \frac{1}{\sin\big(R(T)\big)}.
\]
Recall that $R(T) \to R_* = T_* + \tfrac{\pi}{2}$ as $T \to T_*$; hence $R_* \in (0, \pi)$ and so $\sin(R_*) > 0$. By continuity, there are numbers $\e, \delta > 0$ such that $\sin(R_*) -\e > 0$ and  $1/\sin\big(R(T)\big) < 1/\big(\sin(R_*) - \e\big)$ for all $T \in (T_*, T_* + \delta)$. Therefore $1 / \sin\big(R(T)\big)$ is integrable in a right-sided neighborhood of $T_*$.
\end{itemize}

\item[(b)] Recall $D$ is given by \eqref{eq: U}. Let $R_i = \lim_{t \to - \infty}R\circ \g_i(t)$. Since $R_i = T_i + \tfrac{\pi}{2}$, the endpoints $(T_i, R_i, \omega_i)$ lie on the boundary of $D$. Then (i) - (iii) follow from the geometry of the Einstein static universe. \qed

\end{itemize}

\subsection{The geometric obstruction}

Only properties (i) and (ii) of Theorem \ref{thm: main} were used to establish Proposition \ref{prop: TIFs}.
The next proposition makes use of (iii) and highlights the main geometric obstruction to $C^0$-extendibility. It's an adaptation of \cite[Prop. 3.13]{Sbierski_C0_FLRW} to the spatially flat setting.

\medskip

\begin{prop}\label{prop: obstr}
Let $(M,g)$ be as in Theorem \emph{\ref{thm: main}} and define $(M_1, g_1)$ by \eqref{eq: M_1}. Let $\g_1$ and $\g_2$ be two past-directed past-inextendible timelike curves parameterized by $t$. Let $q \in M_1$ be a point such that the images of $\gamma_1$ and $\gamma_2$ are contained in $M_1 \setminus J^-(q,M_1)$. Let $\omega$ denote the angular coordinates on $M_1$ in spherical normal coordinates. Suppose $\omega_1 = \lim_{t \to -\infty}\omega \circ \gamma_1(t)$ exists and similarly for $\omega_2$. If $\omega_1 \neq \omega_2$, then
\[
d_{\Sigma_t \setminus J^-(q,M_1)}\big(\g_1(t), \g_2(t)\big) \to \infty \quad \text{ as } \quad t \to -\infty,
\]
where $\Sigma_t$ is the spacelike hypersurface in $M_1$ of constant $t$ (in comoving coordinates) and the distance is with respect to the induced Riemannian metric on $\Sigma_t \setminus J^-(q,M_1)$.
\end{prop}

\proof
In spherical normal coordinates, the metric is  $g = -dt^2 + a(t)^2(dr^2 + r^2 d\Omega^2)$.
Using the property that there are no particle horizons, there is a point $p \in J^-(q,M_1)$ such that $r(p) = 0$. (This can be proven analytically but it's easy to see geometrically since $M_1$ is conformal to $D$ within the Einstein static universe.) Consequently, $\Sigma_t \setminus J^-(q,M_1) \subset \Sigma_t \setminus J^-(p,M_1)$, and so it suffices to show 
\[
d_{\Sigma_t \setminus J^-(p,M_1)}\big(\g_1(t), \g_2(t)\big) \to \infty \quad \text{ as } \quad t \to -\infty.
\]
Since $\Sigma_t \setminus J^-(p,M_1) = \big\{(t,r,\omega)\mid r > \int_t^{t(p)}\frac{1}{a(s)}ds\big\}$, any spacelike curve joining $\g_1(t)$ to $\g_2(t)$ lying in $\Sigma_t \setminus J^-(p,M_1)$ will have length atleast
\[
\left(a(t)\int_t^{t(p)}\frac{1}{a(s)}ds\right) d_{\mathbb{S}^{n-1}}\big(\omega\circ\g_1(t), \omega\circ \g_2(t)\big).
\]
Since $\omega_1 \neq \omega_2$, the term $d_{\mathbb{S}^{n-1}}\big(\omega\circ\g_1(t), \omega\circ \g_2(t)\big)$ is bounded below by a positive number for all sufficiently negative $t$. Hence condition (iii) in Theorem \ref{thm: main} implies that the above quantity diverges to $\infty$ as $t \to -\infty$. 
\qed

\subsection{Proof of Theorem \ref{thm: main}}

Let $(M,g)$ be as in Theorem \ref{thm: main} and define $(M_1, g_1)$ by \eqref{eq: M_1}. To prove $(M,g)$ is past $C^0$-inextendible, it suffices to show that $(M_1, g_1)$ is past $C^0$-inextendible by the following lemma.

\medskip

\begin{lem}
Let $(M,g)$ be a spacetime. Suppose there is a proper open subset $M' \subset M$ such that for every future-directed and past-inextendible timelike curve $\g \colon (a,b] \to M$, there is a $t \in (a,b)$ such that $\g\big((a,t]\big) \subset M'$. Then $(M, g)$ is past $C^0$-inextendible if $(M',g|_{M'})$ is past $C^0$-inextendible.
\end{lem}

\proof
We prove the contrapositive. Suppose $\iota \colon (M,g) \to (\wt{M}, \wt{g})$ is a $C^0$ extension with $\pd^-\iota(M) \neq \emptyset$. By definition there is a $\wt{p} \in \wt{M} \setminus \iota(M)$ and a timelike curve $\wt{\g} \colon [0,1] \to \wt{M}$ with $\wt{\g}(0) = \wt{p}$ such that $\g := \iota^{-1} \circ \wt{\g}|_{(0,1]}$ is a future-directed past-inextendible timelike curve in $M$. Define $\iota' \colon (M', g|_{M'}) \to (\wt{M}, \wt{g})$ via $\iota' = \iota|_{M'}$. By assumption, there is a $t \in (0,1)$ such that $\g\big((0,t]\big) \subset M'$. Therefore $\wt{p} \in \pd^-\iota'(M')$.
\qed

\medskip
\medskip

\noindent\underline{\emph{Proof that $(M_1, g_1)$ is past $C^0$-inextendible}}:

\smallskip

The proof follows the same strategy as the proof of \cite[Thm. 1.6]{Sbierski_C0_FLRW}.

Seeking a contradiction, suppose there is a continuous extension $\iota \colon (M_1,g_1) \to (\wt{M}, \wt{g})$ such that $\pd^-\iota(M_1) \neq \emptyset$. For notational simplicity, we identify $M_1$ with its image $\iota(M_1)$, viewing $M_1$ as a proper open submanifold of $\wt{M}$. 

There is a point $p \in \pd^- M_1$. Using global hyperbolicity of $(M_1, g_1)$, along with the assumption that the spacetime dimension satisfies $n + 1 \geq 3$, we can apply the time-reversed version of \cite[Prop. 2.5]{Sbierski_C0_FLRW} to produce a past boundary chart $\phi = (x^0, x^1, \dotsc, x^n) \colon V \to (-\e_0, \e_0) \times (-\e_1, \e_1)^n$ centered around $p$ such that the graphing function $f\colon (-\e_1, \e_1)^n \to (-\e_0, \e_0)$ is differentiable at the origin and $\pd_1f(0) = 0$; hence the graph of $f$ has a spacelike direction at the origin. Recall that each point on the graph of $f$ is a point belonging to $\pd^- M_1$ and everything above the graph of $f$ lies in $M_1$, i.e., $V_> := \phi^{-1} \circ \{x^0 > f(x^1, \dotsc, x^n)\}\subset M_1$. Moreover, the graph of $f$ is achronal with respect to smooth future-directed timelike curves in $V$, where the time-orientation on $V$ is determined by $\pd_0$. Moreover, we can assume that $\wt{g}$ on $V$ satisfies $g_n \prec \wt{g} \prec g_w$, where 
\begin{align*}
 g_n &:= -\tfrac{1}{2} (dx^0)^2 + (dx^1)^2 + \dotsb +(dx^n)^2,
 \\
 g_w &:= -2(dx^0)^2 + (dx^1)^2 + \dotsb +(dx^n)^2.
\end{align*}
Here $\wt{g} \prec g_w$ means $\wt{g}(X,X) \leq 0$ implies $g_w(X,X) < 0$ for nonzero vectors $X$. 

For convenience, we work within the following globally hyperbolic subset $U \subset V$ defined as follows: Let $\e > 0$. Set $a,b \in V$ by $\phi(a) = (-\e, 0, \dotsc, 0)$ and $\phi(b) = (\e, 0, \dotsc, 0)$. Choose $\e$ small enough so that 
\[
U := I^+_{g_w}(a,V) \cap I^-_{g_w}(b,V)
\]
compact closure within $V$. Then $U$ admits a Cauchy surface, $U \cap \phi^{-1}\big(\{x^0 = 0\}\big)$; hence the causal relation $J^{\pm}$ is closed within $U$ \cite[Prop. 3.21]{Ling_C0_causal}.  Moreover, for any two points $x,y \in U$ with $y \in I^+(x,U)$, we have $J^+_{g_w}(x,U) \cap J^-_{g_w}(y,U)$ is compact within $U$. 
Set $U_> := V_> \cap U$, i.e., $U_>$ denotes those points in $U$ that lie above the graph of $f$.

Fix a point $s \in U$ lying on the positive $x^0$-axis. Let $q \in U$ be a point on the graph of $f$ such that $0 < x^1(q) \ll x^0(s)$ and $x^i(q) = 0$ for $i = 2, \dotsc, n$.  Define the curves $\g_1, \g_2 \colon [0,1] \to U$  in coordinates $x^\mu$ by $\g^\mu_1(\tau) = p^\mu + \tau(s^\mu - p^\mu)$ and $\g^\mu_2(\tau) = q^\mu + \tau(s^\mu - q^\mu)$. 
 Choose $x^1(q)$ small enough so that $\g_2$ is timelike. Therefore $\g_1|_{(0,1]}$ and $\g_2|_{(0,1]}$ are future-directed and past-inextendible timelike curves in $M_1$ with future endpoint $s$. Note that $\g_1$ lies on the $x^0$-axis and $\g_2$ lies in the $(x^0,x^1)$-plane. Moreover, since $\pd_1f(0) = 0$, we can choose $q$ close enough to $p$ to ensure that $I^+(p,U)$ is not contained in $I^+(q,U)$ nor vice versa. Lastly, choose $x^1(q)$ small enough so that 
\[
\phi^{-1}\left(\big[0, x^0(s)\big] \times \big[0, x^1(q)\big]^n\right) \subset U \quad \text { and } \quad \phi^{-1}\left(\{x^0(s)\} \times \big[0, x^1(q)\big]^n\right) \subset I^+_{g_n}(s', U),
\] 
where $s'$ is the point $x^0$-axis given by $x^0(s') = \tfrac{1}{2}x^0(s)$.

With $s$ and $q$ fixed, choose $r \in U_>$  such that $x^1(r) < 0$ and $x^i(r) = 0$ for $i =2,\dotsc, n$. Choose $|x^1(r)|$ small enough so that $r \in I^-(s,U)$. It follows from the achronality of $f$ that $r \in I^-(s,M_1)$. Since $\pd_1f(0) =  0$, we can also ensure that $J^-(r,U_>)$ does not intersect the sets 
\[
J^+_{g_w}(p,U), \quad J^+_{g_w}(q,U), \quad \text{ and } \quad \phi^{-1}\big(\{x^1 \geq 0\}\big),
\]
by choosing $r$ to lie just slightly above the graph of $f$.

We claim that the following two properties hold.

\begin{itemize}
\item[(1)] $I^+(p, U) \cap I^-(s, U)$ and $I^+(q, U) \cap I^-(s, U)$ are nonempty and neither set is contained in the other.
\item[(2)] $\g_1|_{(0,1]}$ and $\g_2|_{(0,1]}$ are both contained in $M_1 \setminus J^-(r,M_1)$.
\end{itemize}

(1) follows easily since $I^+(p,U)$ is not contained in $I^+(q,U)$ nor vice versa. However, (2) is not immediate. To prove (2), note that Proposition 3.11 in \cite{Sbierski_C0_FLRW} generalizes to the spatially flat setting. (Here the simple connectivity of $M_1$ is used.) From this we conclude that no past-directed null geodesic in $M_1$ emanating from $r$ intersects one emanating from $s$. Now apply the time-reversed version of Proposition 2.6 in \cite{Sbierski_C0_FLRW} (with $p$ playing the role of $\tilde{t}$). Therefore 
\[
J^-(r,M_1) \cap \bigg(\big[J^-(s,U_>)\cap J^+_{g_w}(p,U)\big]\setminus J^-(r,U_>) \bigg)
=\emptyset.\]
However, since $J^-(r,U_>) \cap J^+_{g_w}(p,U) = \emptyset$, we have  
\[
J^-(r,M_1) \cap \big[J^-(s,U_>)\cap J^+_{g_w}(p,U)\big]
=\emptyset.
\]
It follows that $\gamma_1|_{(0,1]}$ is contained in $M_1 \setminus J^-(r,M_1)$. 
 Replacing $p$ with $q$ in the above argument shows that the same holds for $\gamma_2|_{(0,1]}$. Thus (2) holds.

$(M_1, g_1)$ is future one-connected; see the proof of \cite[Prop. 3.1]{Sbierski_C0_FLRW} or \cite[Prop. 2.7]{Gal_Ling_C0_remarks}).  (Here the simple connectivity of $M_1$ plays an essential role.) Thus we can infer from 
\cite[Lem. 2.12]{Sbierski_holonomy} that
\[
\bigcup_{0 < \tau < 1} I^+\big(\g_1(\tau), M_1) \cap I^-(s,M_1)  \,=\, I^+(p,U) \cap I^-(s,U)
\]
and
\[
\bigcup_{0 < \tau < 1} I^+\big(\g_2(\tau), M_1) \cap I^-(s,M_1)  \,=\, I^+(q,U) \cap I^-(s,U).
\]
By (1),  neither $\bigcup_{0 < \tau < 1} I^+\big(\gamma_1(\tau), M_1)$ nor $\bigcup_{0 < \tau < 1} I^+\big(\gamma_2(\tau), M_1)$ is contained in the other. Set $T_i := \lim_{\tau \to 0} T \circ \g_i(\tau)$ and $\omega_i := \lim_{\tau \to 0} \omega \circ \gamma_i(\tau)$, for $i = 1,2$. Part (ii) of Proposition \ref{prop: TIFs}(a) implies that $T_1,T_2 \in (-\tfrac{\pi}{2}, \tfrac{\pi}{2})$, while parts (i) and (ii) of Proposition \ref{prop: TIFs}(b) imply that $\omega_1 \neq \omega_2$.  Then property (2) above puts us in the setting of Proposition \ref{prop: obstr} (with the point $r$ playing the role of $q$ in the statement of Proposition \ref{prop: obstr}).

Now we establish the contradiction to the conclusion of Proposition \ref{prop: obstr}.  Recall $\Sigma_t$ denotes the smooth spacelike Cauchy surface in $M_1$ of constant $t$. Since the Cauchy surfaces foliate $M_1$, there is a $t_0 > -\infty$ such that $s' \in I^+(\Sigma_t, M_1)$ for all $-\infty < t \leq t_0$. Let $z$ be a point in $M_1$ which corresponds to a point in $\{x^0(s)\} \times \big[0, x^1(q)\big]^n$. From $z$, we can generate a past-directed past-inextendible timelike curve (by moving straight down the $x^0$-axis) which must intersect each $\Sigma_t$ at a unique point. Thus, for $-\infty < t \leq t_0$, we can graph this portion of $\Sigma_t$ via a smooth function 
\[
h_t \colon \big[0, x^1(q)\big]^n \to (-\e_0, \e_0).
\]
The timelike curve $\g_1$ intersects the graph of $h_t$ on the $x^0$-axis at $h_t(0, \dotsc, 0)$, while $\g_2$ intersects the graph of $h_t$ at some $x^1$-value $0 < \a_t < x^1(q)$ in the $(x^0,x^1)$-plane. Connect these two intersections by the smooth curve $\s_t \colon [0,1] \to \S_t \cap U_>$ defined by
\[
\phi \circ \s_t(\tau) \,=\, \big(h_t(\tau \a_t, 0, \dotsc, 0), \tau  \a_t, 0, \dotsc, 0\big).
\]
Since $U$ is locally Minkowskian (in particular $g_n \prec \wt{g}$), we have $|\pd_i h_t| < \sqrt{2}$ uniformly in $\big[0, x^1(q)\big]^n$ and for all $-\infty < t \leq t_0$. Consequently, the lengths $L_{\wt{g}}(\s_t)$ of the spacelike curves $\sigma_t$ are bounded above for all $-\infty <t \leq t_0$. This immediately contradicts Proposition \ref{prop: obstr} provided  $\s_t$ does not intersect $J^-(r,M_1)$ for all sufficiently negative values of $t$.

To show this, recognize that the convergence $h_t \to f$ as $t \to -\infty$ is pointwise monotone by achronality of the Cauchy surfaces; hence it is uniform by Dini's theorem. Therefore for all $\epsilon > 0$ there is a $t_{\epsilon}$ such that for all $t \in (-\infty, t_\epsilon)$, the curve $\sigma_t$ lies in the region of the $(x^0,x^1)$-plane defined by $0 \leq x^1 \leq x^1(q)$ and $0 < x^0 - f(x^1, 0, \dotsc, 0) < \epsilon$. Choose $\epsilon$ small enough, so that the family $\{\sigma_t\}_{-\infty < t < t_\epsilon}$ is contained\footnote{An argument for this containment without appealing to  Dini's theorem can be made with an application of the mean value theorem together with the bound $|\pd_ih_t| < \sqrt{2}$.} in
$J^-(s, U_>) \cap J^+_{g_w}(-s, U)$, where $-s$ is the point on the negative $x^0$-axis given by $x^0(-s) = -x^0(s)$. Moreover, since $r$ was chosen such that $J^-(r,U_>)$ does not intersect the set $\{x^1 \geq 0\}$, it follows that 
\[
\{\sigma_t\}_{-\infty < t < t_\epsilon} \subset \big[J^-(s,U_>)\cap J^+_{g_w}(-s,U)\big]\setminus J^-(r,U_>).
\]
But then another application of the time-reversed version of Proposition 2.6 in \cite{Sbierski_C0_FLRW} (with $-s$ playing the role of $\tilde{t}$) gives
\[
J^-(r,M_1) \cap \bigg(\big[J^-(s,U_>)\cap J^+_{g_w}(-s,U)\big]\setminus J^-(r,U_>) \bigg)
=\emptyset.\]
Thus $\{\sigma_t\}_{-\infty < t < t_\epsilon}$ does not intersect $J^-(r,M_1)$. 
\qed

\section*{Acknowledgments} 
The author thanks Jan Sbierski for helpful comments and discussions. This work was supported by Carlsberg Foundation CF21-0680, Danmarks Grundforskningsfond CPH-GEOTOP-DNRF151, and the Austrian Science Fund (FWF) [Grant DOI 10.55776/EFP6].

%
%
%
%

\let\oldbibliography\thebibliography
\renewcommand{\thebibliography}[1]{
  \oldbibliography{#1}
  \setlength{\itemsep}{1pt}
}



\end{document}